\documentclass[conference]{IEEEtran}
\IEEEoverridecommandlockouts
\usepackage{cite}
\usepackage{amsmath,amssymb,amsfonts}
\usepackage{algorithmic}
\usepackage{graphicx}
\usepackage{subcaption}
\usepackage{textcomp}
\usepackage{multicol}
\usepackage[section]{placeins}
\usepackage{float}

\def\BibTeX{{\rm B\kern-.05em{\sc i\kern-.025em b}\kern-.08em
    T\kern-.1667em\lower.7ex\hbox{E}\kern-.125emX}}
\begin{document}

\title{Detecting Community Structure in Dynamic Social Networks Using the Concept of Leadership}

\author{\IEEEauthorblockN{Saeed Haji Seyed Javadi}
\IEEEauthorblockA{\textit{School of Computing and Information} \\
\textit{University of Pittsburgh}\\
seh138@pitt.edu}
\and
\IEEEauthorblockN{Pedram Gharani}
\IEEEauthorblockA{\textit{School of Computing and Information} \\
\textit{University of Pittsburgh}\\
peg25@pitt.edu}
\and
\IEEEauthorblockN{Shahram Khadivi}
\IEEEauthorblockA{\textit{Department of Computer Engineering} \\
\textit{Amirkabir University of Technology}\\
khadivi@aut.ac.ir}

}

\maketitle

\begin{abstract}
Detecting community structure in social networks is a fundamental problem empowering us to identify groups of actors with similar interests. There have been extensive works focusing on finding communities in static networks, however, in reality, due to dynamic nature of social networks, they are evolving continuously. Ignoring the dynamic aspect of social networks, neither allows us to capture evolutionary behavior of the network nor to predict the future status of individuals. Aside from being dynamic, another significant characteristic of real-world social networks is the presence of leaders, i.e. nodes with high degree centrality having a high attraction to absorb other members and hence to form a local community. In this paper, we devised an efficient method to incrementally detect communities in highly dynamic social networks using the intuitive idea of importance and persistence of community leaders over time. Our proposed method is able to find new communities based on the previous structure of the network without recomputing them from scratch. This unique feature, enables us to efficiently detect and track communities over time rapidly. Experimental results on the synthetic and real-world social networks demonstrate that our method is both effective and efficient in discovering communities in dynamic social networks.
\end{abstract}

\section{Introduction}
The advent and growing popularity of online social networks such as Facebook, LinkedIn, and Twitter had a significant impact on the study of social networks. One of the most important topics in the social network analysis is the problem of finding latent communities. A community in a network is a set of nodes that are densely interconnected to one another while loosely connected to the rest of the network \cite{b1}. One could analyze and understand the structures and functions of a complex network profoundly by detecting the communities. This principal problem has been studied heavily in the past decade. A large number of fast and accurate methods have been developed by researchers in various fields of study \cite{b2}. One of the premier measurement functions called modularity was introduced by Newman et al. to evaluate the quality of detected structures in communities \cite{b3}. The concept of modularity let a new category of methods emerged to detect densely connected nodes in complex networks \cite{b4}–--\cite{b6}. Their strategy was to find a good clustering by maximizing the modularity function. Since maximizing the modularity pertains to the class of NP-complete problems \cite{b7}, several heuristic approaches were proposed to find the near optimal community structure \cite{b5}, \cite{b8}. Finding the local community of a given node is another strategy in community mining which has showed its efficiency in the face of very large complex networks like World Wide Web \cite{b2}. The community is usually formed by expanding from an initial "seed" node as long as the defined local metric strictly improves \cite{b9}–--\cite{b11}. Most of the seed-centric community detection solutions are sensitive to the position of initial source nodes. As forming local cluster around a low degree node usually results in poor quality.
     There have been extensive works focusing on finding communities in static networks \cite{b2}, however, until recently, most of the proposed algorithms ignore the dynamic aspects of social networks by discarding time information of interactions. In the static approach, a social network is treated as a single constant graph that is mostly derived by aggregation of the whole network over time. However, in reality, due to the dynamic nature of the social networks, they continuously evolve. These changes could be joining (leaving) actors to (from) the network and establishing new connections or destroying the existing ones \cite{b12}. This makes a highly dynamic network which witnesses a wide variety of changes. Examples include online social networks such as Facebook \cite{b13}, email exchange networks \cite{b14}, blogosphere \cite{b15} and etc. Since these methods ignore dynamic information associated with ever-changing social networks, they can neither capture the evolutionary behavior of the network nor predict future status of community structure.
     Over recent years, there has been a new trend in devising efficient algorithms to detect communities as well as tracking them in dynamic social networks. A dynamic network can be modeled as a sequence of static networks called graph snapshots where each snapshot corresponds to a particular timestep.
     As we will discuss in more detail in a related work, some methods employ a two-stage mining approach in which at first, a static clustering method is applied on all snapshots and then, obtained communities will be compared with one another to track evolution of community structure over time. Since computing communities is usually independent of the past history, detected community structure of every certain snapshot is dramatically different from the ones related to the other snapshots, especially in noisy datasets. Another type of methods attempts to find a good clustering for each snapshot whereas the detected partition is not different from its history. This could be done by optimizing an objective function composed of two variables named community quality and community history. The main drawback of this approach is that it is not parameter-free. For instance, the method proposed in \cite{b16} requires the number of desired communities which is usually unknown in practice.
     Considering a community as an evolving structure, it can be detected by an incremental updating. The strategy is to keep the community structures of previous steps, unless any changes occur in the underlying network, that is to say whenever a new link is added or an existing one is removed, an update procedure is applied to adapt the clustering to the new structure. It has been demonstrated that this strategy is fast and it keeps community smoothness over time \cite{b17}, \cite{b18}. Yet, highly changing networks could compromise the quality of the result of an incremental community detection algorithm.
     The problem of finding communities in an evolving network is addressed in the literature of multiplex network analysis as well. A multiplex network is defined as a set of networks linked through interconnected layers. Each layer is composed of the same set of nodes which may be interconnected with different types of links. Multilayer networks can also be used to model dynamic networks. In \cite{b19} authors introduced a multi-slice generalization of modularity measure to quantify the quality of the detected community structure in a multiplex network.
    Apart from being dynamic, another significant characteristic of real-world social networks is the presence of leaders, i.e. influential members in local communities, which is an old topic in the field of social science \cite{b20}. Until now, finding leaders and analyzing their social influences in various types of social networks is still an appealing topic \cite{b21}--\cite{b23}. In regard to this important property, recently there has been a new trend of community detection methods by means of community leaders as the pivotal members. The central position of a leader makes it a good option to be chosen as the initial source node in the seed-centric local community detection methods. Hence, different definitions of centrality were applied to distinguish desirable leaders from non-leader members \cite{b24}.
     Following this idea, we designed an efficient method to incrementally detect the communities in the dynamic social networks using the intuitive idea of importance and persistence of community leaders over time. Briefly, in the proposed method, community leaders of each timestep are detected efficiently and the community membership of next timestep is determined by using these influential members as the initial seed nodes. Our proposed method has following key properties:
\begin{itemize}

\item \textbf{Online:} in order to find the meaningful communities at each timestep, the structural information of the previous snapshot is employed.
\item \textbf{Fast:} since our algorithm employed an incremental fashion to the evolving communities, it does not need to calculate the communities of each snapshot from scratch, the runtime of the method is reduced considerably.
\item \textbf{Smooth:} in real-world social networks, community memberships are not expected to change abruptly, so the detected community structure of adjacent graph snapshots should not differ a lot.
\item \textbf{Parameter-free:} unlike most of the solutions proposed in the literature, our algorithm does not require any parameter settings.
\item \textbf{Adaptive:} unlike most incremental methods, this algorithm is efficient even for highly dynamic networks.
\end{itemize}
     The remainder of this paper is organized as follows: In the following section we review the existing approaches related to our work. In section 3 after the preliminaries, the problem definition is provided. Then, we present the proposed method in more details. Section 4 is dedicated to evaluate the method by comprehensive experiments on several real-world and synthetic dynamic networks. Finally, we present conclusions and future directions of our research in section 5.

\section{Related Work}
In this section, a survey on some existing works related to the contributions of the proposed method is provided. First, we review several notable seed-centric methods and then, we discuss three main existing approaches to detect the community structure in the dynamic social networks.

\subsection{Seed-Centric approach}
In addition to a vast number of designed algorithms to unfold the community structure of a complex network by maximizing a global fitness function like modularity, an alternative approach is introduced recently to detect the communities. The main idea of this approach, which is called seed-centric, is to determine local community around a given node. At first, it was accepted as a suitable approach for finding local community structure; subsequently, several local quality functions were introduced in \cite{b9}–--\cite{b11} to measure the goodness of the detected local clusters. The authors also devised greedy methods to expand a local community around the source node by maximizing their quality measure. Since these methods are sensitive to the position of initial seed nodes, Chen et al. \cite{b25} chose the local maximal degree as the seed of their algorithm and formed the local cluster by iteratively adding neighbor nodes to it. To add suitable adjacent nodes to the core of the local communities, they have tested multiple quality functions. Although they reached higher accuracy compared with the other local community detection methods, but due to high time complexity of their approach, it is not scalable in finding communities in large social networks.
     Seed-centric approach is not limited to the local community detection problem; DOCNet is another efficient algorithm to identify overlapping communities on the entire network. The authors used two concepts of compactness and separability \cite{b26} to create new objective function called index of connectivity. Their main strategy is to select a central influential node and add proper nodes to expand the community until a stopping criterion is met \cite{b27}. Since in these works, the initial seed nodes are high central members, they usually referred as community core or community leaders \cite{b24}. With the assumption that every group of individuals in communities is composed of two types of members: leaders and followers, Rabbany et al.  \cite{b28} found the k-most central nodes as top leaders which could be followed by non-leader nodes to form communities. They examined a wide variety of central measures to find the most appropriate definition of leaders. However, their method is not parameter-free and requires the number of the desired communities, which most of the time is unknown in advance. In \cite{b29} the characteristics of major seed-centric methods are discussed in more depth.

\subsection{Community detection in dynamic networks}

Usually, dynamic networks are represented as a series of static graphs over a period of discrete timesteps, called snapshots. Each snapshot corresponds to a particular timestep of the dynamic network. A community in a dynamic social network is not only a group of densely interconnected nodes, but its members keep their close interactions over an expected time. This problem is also known as dynamic graph clustering. Most of the proposed methods to detect communities in temporal networks could be categorized into three main strategies:   
    The first strategy is a two-stage method (also known as independent clustering) which is based on slicing the whole network as a series of snapshots. Its basic idea is to apply a static clustering method to each snapshot independently and capture the evolution of the communities by comparing the clustering of the consecutive snapshots. Based on this approach, Hopcroft et al.  \cite{b30} were of the first authors who applied a static clustering on each snapshots and they tracked communities over time by the clusters’ similarities over consecutive timesteps. They showed that even small perturbations in the graph could lead significant changes in the structure of the detected communities. This is the main disadvantage of two-stage methods, they are vulnerable to even small changes in network structures and so the output of these methods would result in noisy and short-life communities. To resolve this problem, the authors introduced the natural communities which could remain stable under several perturbations in a graph structure. Based on the same consideration and with a different methodology, Seifi and Guillaume \cite{b31} tracked the evolution of community cores instead of all members of communities. They defined community cores as a set of nodes that frequently clustered together during several executions of a non-deterministic community detection method. They showed that in a dynamic social network, this group of nodes is much more stable compared with all members of a community.
     Although core-based methods could reduce the variance caused by unstable nodes, but they usually require parameters to specify the borderline of core and non-core members. Moreover, applying several perturbations and applying clustering method on each snapshot to detect significant clusters are very time-consuming. The other drawback of independent clustering is the computing similarities between huge numbers of the communities across multiple snapshots. Since the number of the found communities of each snapshot could be more than the number of nodes, it could be impractical in facing big dynamic networks like the method introduced in \cite{b32}.
     Apart from independent clustering, evolutionary clustering was introduced. In this approach, by using the structural information of the preceding timesteps, the detected communities of the consecutive snapshots should not dramatically differ. In doing so, these methods try to optimize an objective function composed of two variables: the cost of a good partitioning for all snapshots and the cost of the difference between the structure of the communities and its history. FacetNet is a well-known evolutionary framework to find a soft community structure based on generative models. In their proposed framework, the community structure of the current snapshot is detected by incorporating the current graph structure and historic community evolution patterns \cite{b33}. Recently, Kawadia and Sreenivasan \cite{b34} have proposed a new measure called "Estrangement" to quantify the partition distance between two consecutive timesteps. They describe estrangement as "the fraction of intracommunity edges that become inter-community edges as the network evolves to the subsequent snapshot". Applying this constraint on the distance metric, they found temporal communities in each snapshot by maximizing the modularity.
     Another strategy to unfold communities in temporal networks is incremental community detection, which is based on structural changes of the network such as link insertion and link removal. Methods using this approach, consider the evolution of networks as a series of atomic events which could change community memberships of individuals (therefore, it is known as event-based clustering). This strategy has two main advantages when encountering with any changes in the network structure: First, the running time of the algorithm is decreased significantly because it does not need to apply clustering method from scratch. Second, it preserves community smoothness over time because of its dynamic updating strategy.
     Gorke et al. \cite{b35} introduced dGlobal which is a dynamic version of the CNM method \cite{b36} to maximize modularity at each timestep. . They aimed to save and modify the dendogram of the obtained clusters. Community memberships of affected nodes will be determined simultaneously, proportional to structural changes in the network. Similar to that, Nguyen et al.  \cite{b12} proposed another algorithm to find modular communities in each snapshot. Although their method is faster than the method which clusters each snapshot independently \cite{b4}, ), running their method for a long time on a dynamic network will end up in poor quality results. The method proposed by Takaffoli et al. \cite{b37} intends to find the community structure at any time based on the extracted clusters from the previous timestep. They introduced an adaptive algorithm which like two-stage methods, employs an event-tracking framework. At each timestep, they consider the obtained connected components of the communities in the last snapshot as the initial seeds for the current snapshot. They form communities around the seeds by maximizing the ratio of the average internal degree and minimizing the average external degree of the local cluster. However, they did not describe how their framework handles some certain scenarios. An example is the case in which new emerged nodes do not belong to any existing communities and independently create a new community.
     Event-based clustering implicitly assumes that network connections do not change much over time and even so, they have only small impact on the current community structures. But, in contrast, most of real-world social networks do change dramatically even in a short period of time \cite{b38}, \cite{b39}. In this regard, Barabasi stated that dynamics of many social and economic phenomena are driven by bursty nature of human behavior \cite{b40}. Speaking of dynamic social networks, Wang et al. \cite{b41} through their experiments, showed that a large number of nodes appear in less than tree snapshots. In other words, only a small portion of nodes could remain stable during network changes over time. Recently, Hao Xu et al. \cite{b42} studied the problem of detecting stable community cores in mobile social networks. By assuming that the links existing in the community core remain stable during consecutive snapshots, they summarized the problem of detecting dynamic community structure into analyzing community core evolution over time.
\section{Method Description}
\subsection{The main idea}\label{AA}
A good community structure in a dynamic social network is one in which the members should have a strong structured similarity with each other and they should keep this similarity over time. As a result, any method whose aim is to detect and track communities in dynamic social networks should consider two main characteristics:

\begin{enumerate}
\item Detected communities should be modular at each timestep. In other words, nodes tightly connected to one another, have to be grouped together in the same cluster.
\item 2-Temporal smoothness of clusters over consecutive snapshots should be preserved; that is, in most of the cases, the communities of time t should not sharply differ from the ones of time t-1.
\end{enumerate}

In this section we first present some definitions focusing on the terms that we use through this paper. Then, we explain the existence and importance of community leaders in real-world social networks. The formal definition of a leader is introduced just after the significant role of leaders is expressed. The next section is dedicated to introduce our local clustering method expanding communities around the promising leaders. In the following, we discuss an important phenomenon in dynamic social networks, i.e. the advent of new communities. Finally, our robust incremental model will be described.

\subsection{Notations and Definitions}
We model a dynamic social network as a sequence of snapshots 

$\{G^1,G^2,\dots,G^\Delta\}$, where $G^t=(V^t,E^t)$ denotes a static graph at time point $1\leqslant t<\Delta$,$V^t$ and $E^t$ are the set of nodes and edges of graph  $G^t$, respectively. The degree of node $v^t$ denoted by  which is the number of neighbors of  at time point $t$. The in-community degree of $v^t$ is given by $d_{v^t}^{in}$  denotes the number of links that connect $v^t$  to nodes of the same cluster.
     We use $L^{t-1}=\{L_1^{t-1},L_2^{t-1},\dots,L_j^{t-1}\}$  to represent the set of leaders of $j$   communities in the snapshot of $G^{t-1}$  and $C^t=\{C_1^t,C_2^t,\dots,C_j^t\}$ denotes set of communities that are obtained by expanding around corresponding leaders of the previous snapshot, i.e. $L^{t-1}$. Since some nodes may be remained unassigned to any clusters, the set $P^t=\{C_1^t,C_2^t,\dots,C_j^t,C_{j+1}^t,\dots,C_k^t\}$ is introduced. The set of newcomer communities $\{C_{j+1}^t,\dots,C_k^t\}$  is a subset of $P^t$ which is determined by clustering the leftover nodes. Consequently, the set $P^t=\{C_1^t,C_2^t,\dots,C_k^t\}$  partitions $G^t$ into some non-overlapping communities where $\cup_{1\leqslant i \leqslant k}C_i^t=G$ and $C_i\cap C_j=\emptyset$ $ \forall{i,j}$.

\subsection{Existence of leaders in social networks}
\begin{figure}[htbp]
\begin{center}
\includegraphics[scale=.5]{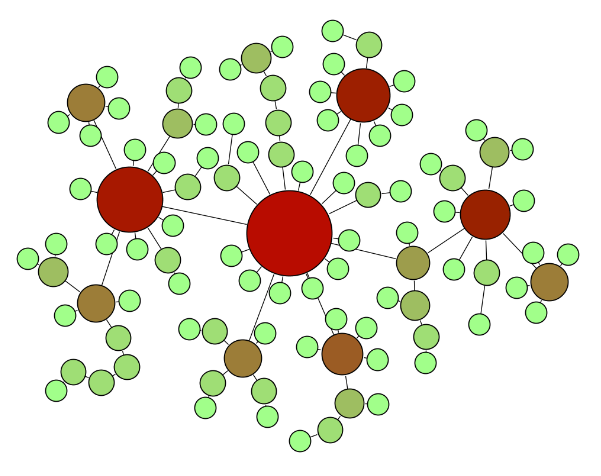}
\end{center}

\caption{A random network generated by the model of Barabasi and Albert.}

\label{fig1}
\end{figure}

Social networks reflect the behavior and interactions of individuals; and as in real-world, some members have great influence on the others. Communities are commonly formed around these influential members called leaders. This observation is one of the results of "preferential attachment" model introduced by Barabasi \cite{b43} ) to explain the power-law distributions of the node degrees in the real-world social networks. In their model, a newcomer node preferentially connects to the nodes that have more connections. These high degree nodes potentially are more popular and have stronger ability to attract new members. This mechanism reproduces the power-law degree distribution observed in the real social networks. When the degree distribution of a graph follows power-laws, it means that a major fraction of the nodes are low-degree and conversely, a small number of them have so many connections (Fig.~\ref{fig1}). This key property has been studied in many real-world networks such as World Wide Web \cite{b44}, citation network \cite{b45} and online social networks \cite{b46}.

     Based on the discussion above, a node with a high degree centrality has a high attraction to absorb other nodes and hence to form a community. This is one of the main characteristics of leaders which could be captured from the local structure of members in a social network. Based on this important feature, it is argued recently for choosing the most central node as the initial seed to detect meaningful local communities. Among dozens of centrality parameters, the maximal-degree nodes are showed to be good choices to select in seed-centric community detection algorithms \cite{b27}, \cite{b28}, \cite{b47}.
     The topology of initial seeds is one of the most important criteria in seed-centric community detection methods. Many of the existing methods expand the community around a single node \cite{b28}, \cite{b47}, whereas choosing a set of central nodes as the initial seeds, leads to less computations and may result in a more robust solution. On the other hand, to ensure obtaining high quality communities, the set of initial nodes should be strictly interconnected to one another.

\subsection{Detecting Community Leaders}

In the proposed method, local communities of each snapshots are determined by initialized seed nodes. So, defining and selecting the leaders have a significant effect on detecting desired clusters. A set of nodes with particular characteristics will be selected as the promising community leaders. These nodes are densely interconnected and form a clique structure. Since a clique is a subgraph whose underlying nodes are fully connected, we could be sure that they would be clustered together when any community detection method is applied. Also, in \cite{b2}, \cite{b48} it is showed that it is better to choose a clique instead of an individual node as an initial seed. 
     Suppose community $C_i^t$ is known, we present the definition of community leader of $C_i^t$ as follows:
Community leaders: Considering node $v$ has the maximum in-community degree of $C_i^t$ , community leaders of this cluster are defined as the intersect of nodes that exists in all maximal cliques containing $v$ .
A maximal clique is a fully connected subgraph which is not a subset of any other cliques. To find maximal cliques in a network, we use the method proposed by Eppstein et al. \cite{b49} which is fast and efficient. Because the method is applied only on the ego network of the node with maximum in-community degree and not on the whole network, we can assure that the total running time of leader detection process would be short.
     In dynamic social networks, we expect the leaders not only to construct the communities, but to be more stable than non-leader nodes over time also. Experimental results on real-world social networks strictly confirm this claim.

\subsection{Local Expansion}\label{SCM}
To associate follower nodes to their correspond leaders, we use "index of connectivity" which is an unbiased local objective function presented by \cite{b27} and it is defined as follows:
\begin{equation}
IC_{C_i^t}=\frac{\eta_{C_i^t}-\mu_{C_i^t}}{\sqrt{{\eta_{C_i^t}+\mu_{C_i^t}}}}
\end{equation}

Where $\eta_{C_i^t}$  represents the number of links connecting members of community $C_i^t$  to one another and  $\mu_{C_i^t}$ is the number of links connecting members of community  $C_i^t$ to the outside of the community. This objective function aims to strengthen internal edges as well as weaken external edges of the cluster. To expand community around the leaders, nodes will be selected that greedily increase index of connectivity. To recalculate the objective function for each candidate node efficiently, we may use the following equation:
\begin{equation}
I{C^\prime}_{C_i^t}=\frac{{\eta^\prime}_{C_i^t}-{\mu^\prime}_{C_i^t}}{\sqrt{{{\eta^\prime}_{C_i^t}+{\mu^\prime}_{C_i^t}}}}
\end{equation}
Where ${\eta^\prime}_{C_i^t}$, ${\mu^\prime}_{C_i^t}$,   and $I{C^\prime}_{C_i^t}$ are corresponding scores after aggregating the node $u$ into $C_i^t$  . Two variables ${\eta^\prime}_{C_i^t}$ and ${\mu^\prime}_{C_i^t}$ are determined as follows:
\begin{equation}
{\eta^\prime}_{C_i^t}=\eta_{C_i^t}+in_{C_i^t}(u)
\end{equation}

\begin{equation}
{\mu^\prime}_{C_i^t}=\mu_{C_i^t}+d(u)-2in_{C_i^t}(u)
\end{equation}
Where $d(u)$  and $in_{C_i^t}(u)$  are the degree and in-community degree of node $u$. We continuously keep expanding the community $C_i^t$ until no further improvement on the objective function could be achieved. 

\subsection{Membership of Nodes}
After all communities are expanded, we would face tree type of nodes: (1) Nodes that are assigned to one community (2) Nodes that are assigned to more than one community and usually sit on the intersections of communities. (3) Nodes that are not assigned to any community because they show low orientation to any existing leader.
In real-world social networks, individuals are often shared between communities \cite{b2}. However, to simplify the model, the most suitable community of each hub will be determined. This is based on the ratio of the intersection of neighbors of   and members of   to the union of them all. Hub   would merge to the community with the highest similarity. The similarity measure is calculated as follows:
     \begin{equation}
     J(u,C_i^t)=\frac{|neighbors(u)\cap members(C_i^t)|}{|neighbors(u)\cup members(C_i^t)|}
     \end{equation}
In case 3 where nodes are not associated to any leader, they may be considered to have their own community. In the next subsection, we will discuss about this common event happening in dynamic social networks.

\subsection{Advent of New Communities}
Through our experiments, we noted that a considerable number of nodes will appear at each new timestep. Usually, the newcomer members are interconnected dense enough to form a new community. Therefore, we expect them not to be associated to any existing community and to remain unassigned. In this case, identified communities are temporarily omitted from the network and a static method is applied to cluster the unassigned nodes. Newly detected clusters are considered as newborn communities. This is an important event that most other incremental methods have failed to handle it properly.

\subsection{Algorithm Description}
Our approach requires an initial set of leaders $L^1$  for the first snapshot. Thus, any static community detection algorithm would be applied on the first snapshot of the network to identify communities of graph  $G^1$. With obtained clusters $C^1$ leaders of corresponding communities could be determined easily by the definition proposed in section 3.4. We represent leaders of the first snapshot by $L^1$. The leaders are the most cohesive structures located around the node of highest in-community degree. Featuring the durability of leaders over time, they could be suitable representatives of the community for the next timestep. Facing a new snapshot, the set of leaders of the current timestep will be employed as the initial seeds for next stage. New communities will locally expand around the seed nodes by employing the procedure discussed in section 3.5.
After local communities are expanded, we choose the most suitable communities for the hub nodes. In addition, it is very likely to have some nodes may belong to no cluster.  In other words, identified cluster $C^t=\{C_1^t,C_2^t,\dots ,C_j^t\}$  which are formed around the   $L^{t-1}=\{L_1^{t-1},L_2^{t-1},\dots ,L_j^{t-1}\}$ do not partition graph  . In this case, identified communities are omitted and a static clustering method like Infomap \cite{b50} is used to cluster the remained nodes. We use ${C^\prime}^t=\{C_{j+1}^t,C_{j+2}^t,\dots ,C_k^t\}$  to denote new clusters detected by the static clustering method. By union of $C^t$ and ${C^\prime}^t$ we partition graph $G^t$  as $P^t=C^t \cup {C^\prime}^t$ . We note that all communities obtained the in second step are considered as newborn communities. Similarly, a community dissolves when all of its corresponding leaders disappear in the next snapshot.

\section{Experiments}

In order to evaluate the performance of the proposed method we conducted comprehensive experiments on both computer generated networks and real-world dynamic ones. Two networks generated by computer and also several real-world dynamic networks were investigated. Studying the persistence of the leaders and non-leader nodes on real-world social network proves that the community leaders of the proposed definition are much more stable in compare with follower nodes. As a baseline for comparison, we applied three well-known community mining algorithms of different approaches. These methods are FacetNet \cite{b16} , Estrangement \cite{b34} and static clustering using FUA method \cite{b4}. The static method does not consider any temporal evolution and is applied on each snapshot independently. We consider running time, community smoothness and similarity of result to ground-truth communities as the evaluation criteria. Before the experiments, both real-world and synthetic datasets are first described.

\subsection{Real-world Dynamic Social Networks}
We tested our method on different real-world dynamic social networks selected from various field of studies such as online social networks, email and cellphone communication networks.
     Catalano cellphone network \cite{b51} consists of communication information of 400 unique telephone numbers for the period of 10 months. The total number of phone calls during this period is 9834. Each corresponding snapshot is created by aggregating all connections during every 24 hours.
     Opsahl network originates from an online community for students at University of California, Irvine \cite{b52}. It consists of 59835 messages sent over 1899 subscribed users from April to October of 2004. Collected information of every two weeks represents a single timestep in our experiments.
     Another network is composed by a set of exchanged emails of the department of computer science at KIT \cite{b53}. This network was collected over 4 years and it is comprised of 1890 users and about 550000 email messages. The corresponding chair ID of each email address is considered as ground-truth community structure. We have sliced this data into snapshots of one week for our experiments.
     Another network contains the wall posts from the Facebook New Orleans networks \cite{b54}. In this dataset, each edge represents a post to other user’s wall and each node represents a Facebook user. This dynamic network was gathered in about 30 months containing 46952 nodes and 876993 links. The time interval between snapshots is selected one month.
     The Enron email network \cite{b55} consists of about 1100000 emails which are sent among 87273  employees of Enron Corporation between beginning of 2001 and 2003, whose graph snapshots are taken monthly.

\subsection{Computer-Generated Networks}
To investigate the accuracy of the proposed algorithm, we tested our method on two artificially generated dynamic networks. In these dynamic graphs, hidden ground-truth communities are embedded in the network data and evolve randomly over time to simulate evolution of real-world communities.

Kawadia and Sreenivasan \cite{b34} introduced a dynamic network generator based on markovian evolution. In their benchmark, a set of consecutive snapshots is generated consisting predefined communities embedded in a random background. The underlying links in the communities undergo markovian evolution to simulate community evolution in real-world scenarios. The initial snapshot is created by Erdos-Renyi random graph model \cite{b56} in which $N$  represents the number of nodes, $p_r$ is the existence probability of any edge in random background. Parameter $p$  indicates the probability of intra-community edges which is independent with $p_c$. An edge disappears from the community by the probability of $p$  and conversely, a new link appears by the probability $q$. To preserve initial edge density within the community, $q$ is set as $q=\frac{pp_c}{(1-p_c)}$.

Considering each community characterized by a series of evolutionary events, Green et al. \cite{b57} proposed synthetic dynamic networks based on LFR benchmark \cite{b58}. They created four networks corresponding four main events occurred during lifetime of a community. These events are as follows:

Intermittent communities: In the corresponding network, 10\% of communities are removed randomly at each timestep.

Expansion and contractions: To examine the effect of rapid expansions and contractions of communities, 40 randomly selected communities, absorb new nodes or lose their former members by 25\% of the previous size. 

Birth and death: At each timestep, 40 new clusters are created by the nodes which have left their former communities. Furthermore, 40 existing communities are removed randomly.

Merging and splitting: In the last case, 40 instances of the existing clusters merged two by two and similarly, 40 communities split into two new communities at each snapshot.

\subsection{Persistence of leaders in real-world networks}
\begin{figure*}[t!]
\begin{multicols}{2}
    \includegraphics[scale=.9]{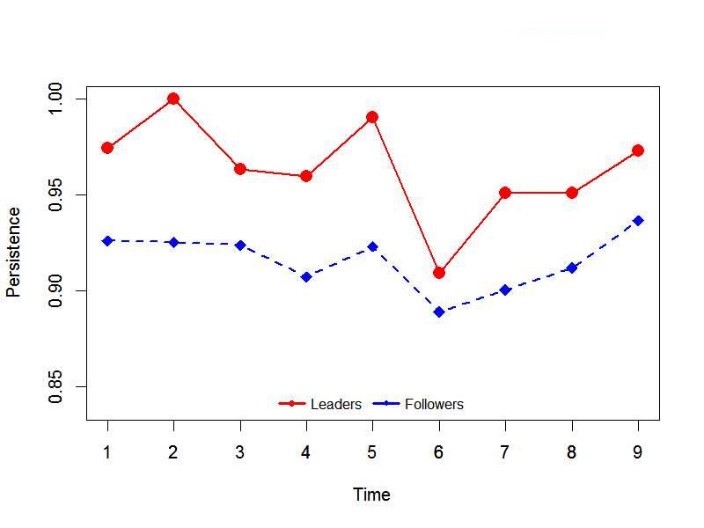}\par 
    \subcaption{Cellphone network}
    \includegraphics[scale=.9]{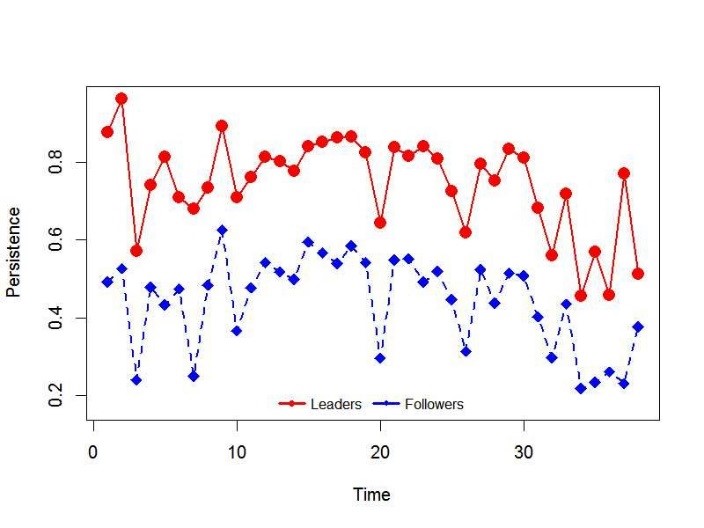}\par 
    \subcaption{Enron network}
    \end{multicols}
\begin{multicols}{2}
    \includegraphics[scale=.9]{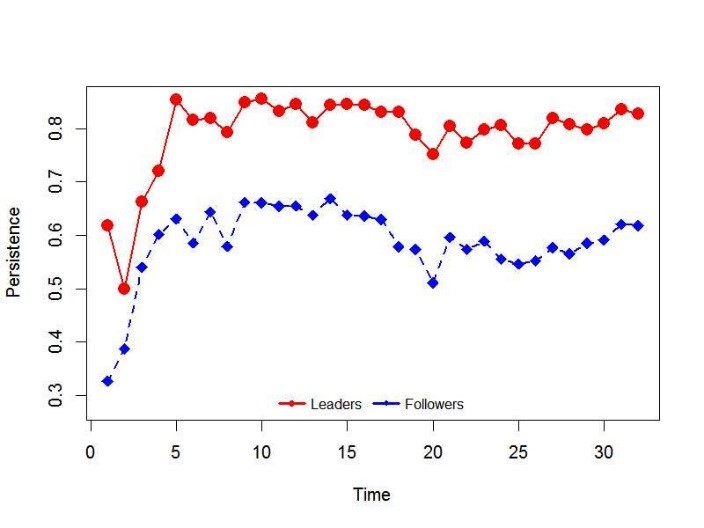}\par
    \subcaption{Facebook network}
    \includegraphics[scale=.9]{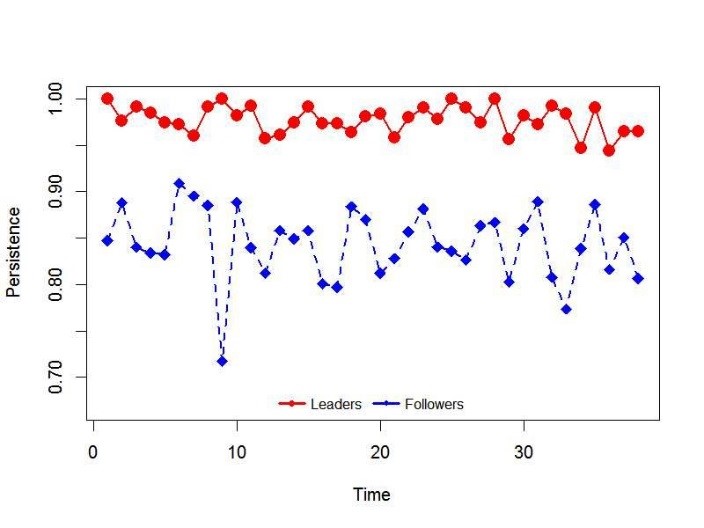}\par
    \subcaption{KIT network}
    
\end{multicols}

	\begin{center}

	\includegraphics[scale=.9]{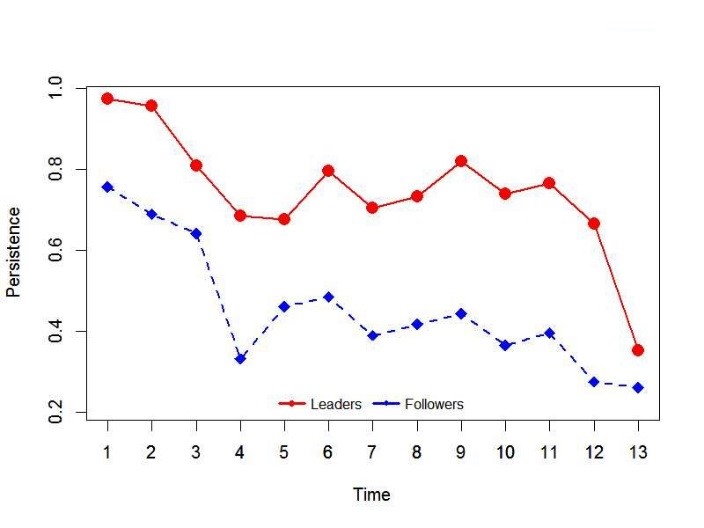}\par
    \subcaption{Opsahl network}
    \end{center}

\caption{Persistenc of leaders and follower on real-world dynamic social networks}
\label{fig2}
\end{figure*}

Before we compare the performance of our algorithm with the baseline methods, we examine the stability of community leaders over time. It is illustrated in Fig.\ref{fig2} that the persistence of leaders versus community followers on all timesteps in social networks introduced. The vertical line of time $t$ shows the ratio of community leaders/followers presented on both $G^t$  and $G^{t+1}$. As it can be seen, the persistence of community leaders is always significantly greater than that of follower nodes. For instance, in time step 10 in KIT network, about 30\% of followers are missed between $G^{10}$ and $G^{11}$, while all community leaders are present on both timesteps.

     Note that, community leaders for each timestep are determined based on the community structure of that snapshot. In other words, the role of community leadership could shift to other members of the group.

\subsection{Temporal smoothness}
In the next experiment we analyze the smoothness of identified communities. Therefore, clustering of successive snapshots is compared together by a similarity measure. Higher similarity value indicates a smoother evolving community structure. To compare the community memberships of two adjacent snapshots, Normalized Mutual Information (NMI) is adopted. Let A and B be two partitions of a same graph, $NMI(A,B)$ equals 1 If A and B are identical and equals 0 if they are totally different. The greater the value of NMI, the more similar the two partitions are.
     In our data sets, some nodes are removed and new members are added over time. As a result, the size of two consecutive snapshots may differ and so NMI cannot be applied. To overcome this issue, we measure the clustering similarity of members attended in both consecutive snapshots.
     We should note that, since FacetNet requires an adjacency matrix for each snapshot to calculate community membership of nodes, its space complexity is more than $\Omega (\Delta n^2)$  . That is, FacetNet is impractical for networks more than 10 thousand nodes. Also, the execution time of Estrangement method is very high, as it does not converge to any solution in 24 hours of runtime. That is why we have omitted the solutions of these two methods for Facebook and Enron datasets. 

\begin{figure*}[h]

\begin{multicols}{2}
    \includegraphics[width=\linewidth]{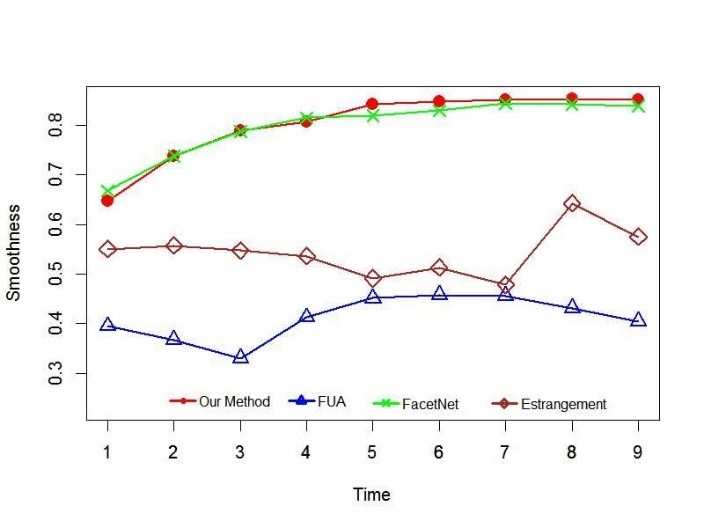}\par 
    \subcaption{Cellphone network}
    \includegraphics[width=\linewidth]{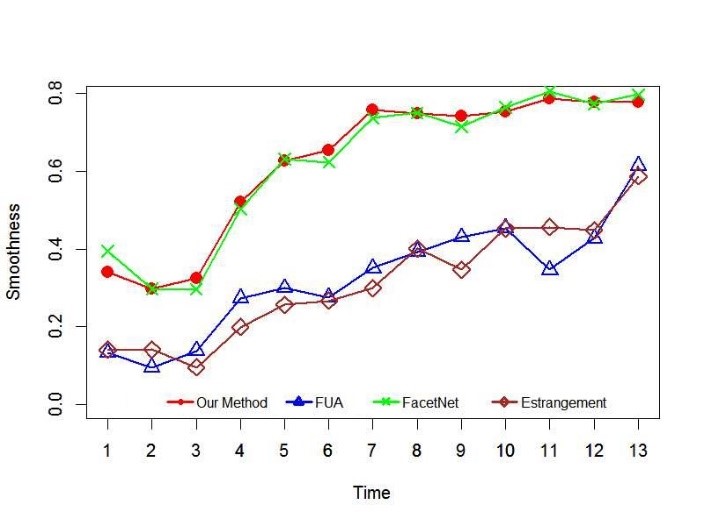}\par 
    \subcaption{Opsahl network}
\end{multicols}

\begin{multicols}{2}    
    \includegraphics[width=\linewidth]{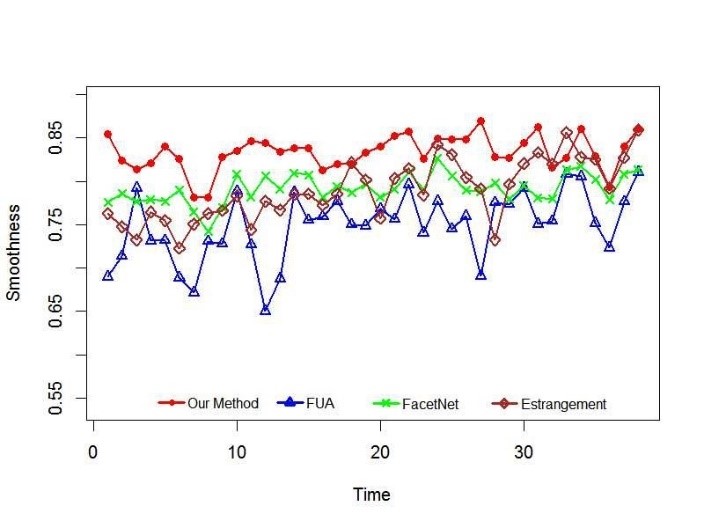}\par
    \subcaption{KIT network}
    \includegraphics[width=\linewidth]{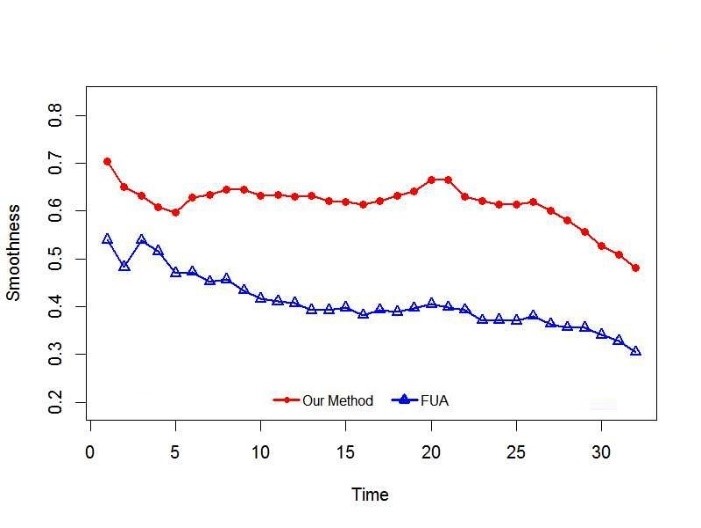}\par
    \subcaption{Facebook network}
\end{multicols}

\begin{center}
    \includegraphics[scale=1]{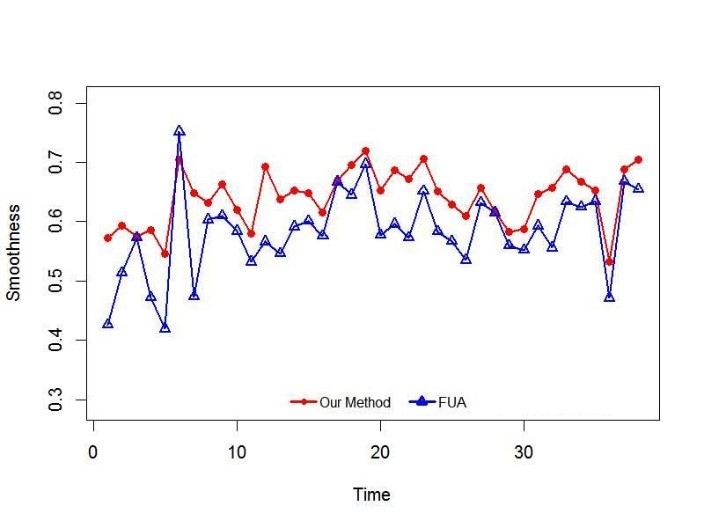}\par
    \subcaption{Temporal smoothness of various methods on Enron}    

\end{center}

\caption{Temporal smoothness of dynamic clustering methods on real-world social network}

\label{fig3}
\end{figure*}

\begin{figure*}[h]
\begin{multicols}{2}

    \includegraphics[width=\linewidth]{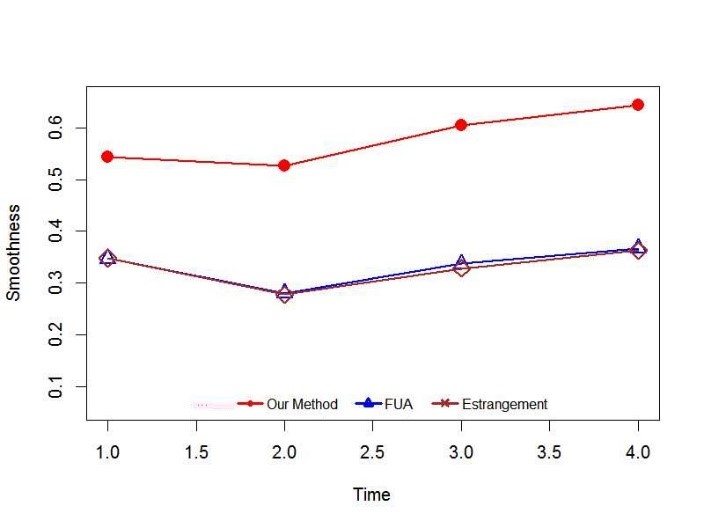}\par 
    \subcaption{"intermittent" communities}
    \includegraphics[width=\linewidth]{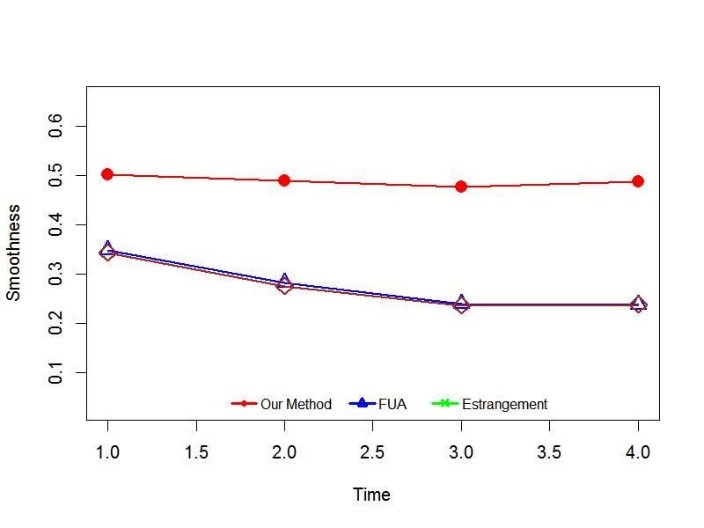}\par 
    \subcaption{Expansion and contraction events}
    
    \end{multicols}
\begin{multicols}{2}
	\includegraphics[width=\linewidth]{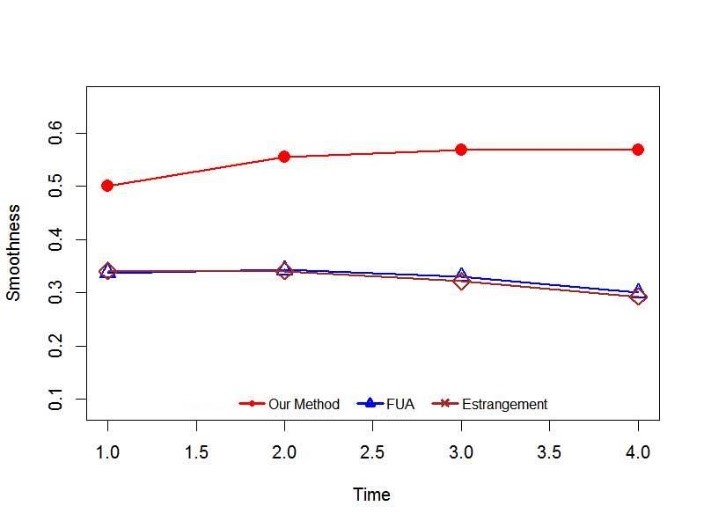}\par
    \subcaption{Birth and death events}
    
    \includegraphics[width=\linewidth]{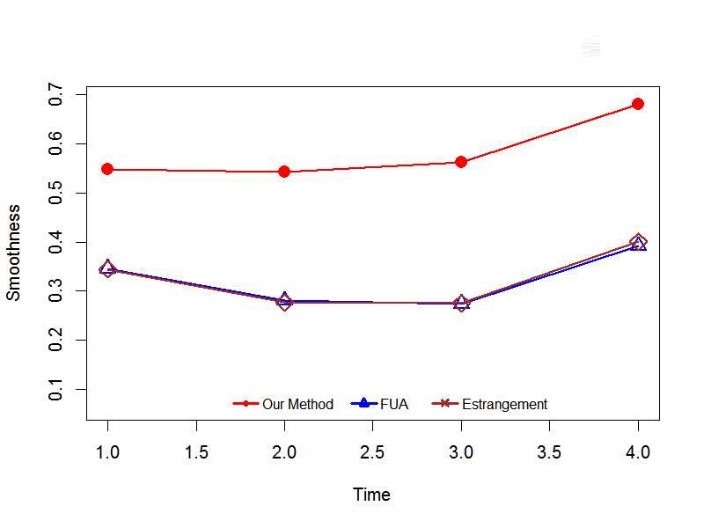}\par
    \subcaption{Merging and splitting events}
        
\end{multicols}
    \begin{center}
    \includegraphics[scale=1]{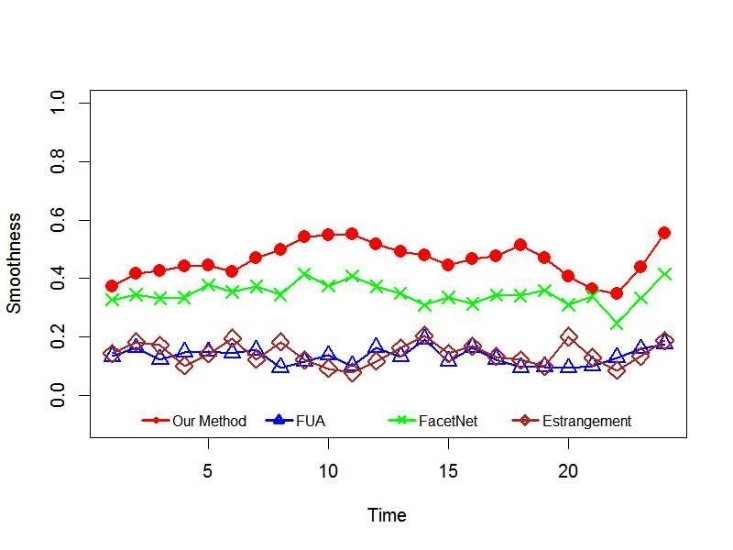} \par
    \subcaption{Kawadia synthetic network with  $n=100$, $P=0.4$, $P_c=0.2$, $P_r=0.05$ and $n_c=20$}
    \end{center}
\caption{Temporal smoothness of dynamic clustering methods  on synthetic datasets}
\label{fig4}
\end{figure*}

    As we can see in Figures \ref{fig3} and \ref{fig4}, our method achieves higher temporal smoothness compared with several baseline methods, except for Opsahl and cellphone datasets in which the proposed algorithm and FacetNet have a close competition. As expected, the independent clustering method ranks last because it does not use any structural information of previous partitions to cluster the current graph snapshot.

\subsection{Similarity to ground-truth communities}
In the next evaluation phase, we investigate the similarity of clusterings to the ground-truth communities. In order to measure the similarity, we employed NMI as well. We compare the performance of proposed method with the other approaches on KIT. Among the available datasets, KIT is the only real-word social network which contains metadata as the ground-truth community structure. 

\begin{figure}[H]
\centering
\includegraphics[scale=.5]{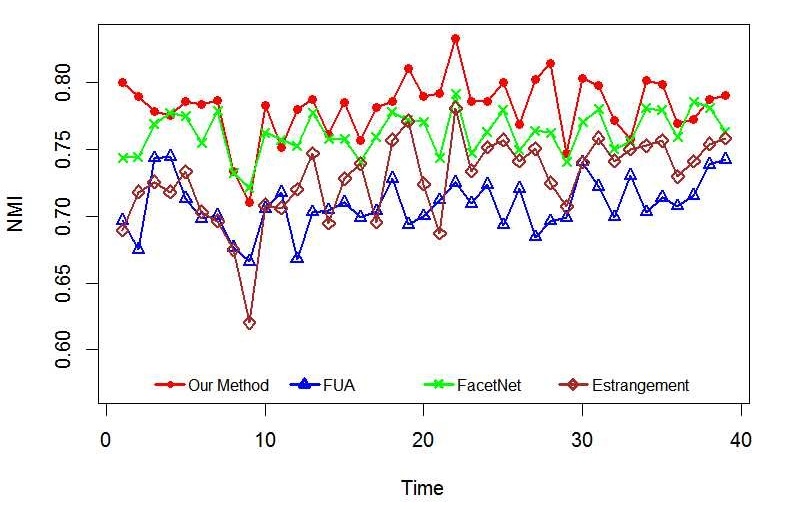}
\caption{Similarity to ground-truth community in KIT network}

\label{fig5}
\end{figure}

The results showed in Fig.\ref{fig5}, confirm that detected communities of the proposed method are the most similar partitions to the known community structure of KIT. Although the FacetNet is the next accurate method, it requires the number of desired communities as input which could result in completely different solutions using different parameter settings. In order to ensure a fair comparison, this parameter is set as the average number of known communities in KIT. However, in practice the true number of clusters is unknown beforehand.
     
     The accuracy of Estrangement method considerably fluctuates over time. But overall, it shows a better performance in compare with FUA method. A possible explanation would be that, Estrangement optimizes modularity function by considering the community structure of previous snapshot, whereas independent clustering aims at maximizing modularity merely which may cause missing some important evolution details.
     
\begin{figure*}[h]
\begin{multicols}{2}
    \includegraphics[width=\linewidth]{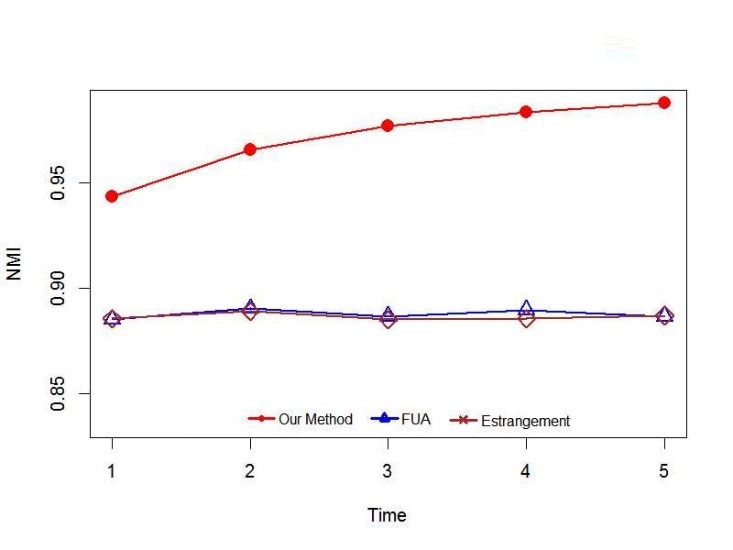}\par 
    \subcaption{"intermittent" communities}
    \includegraphics[width=\linewidth]{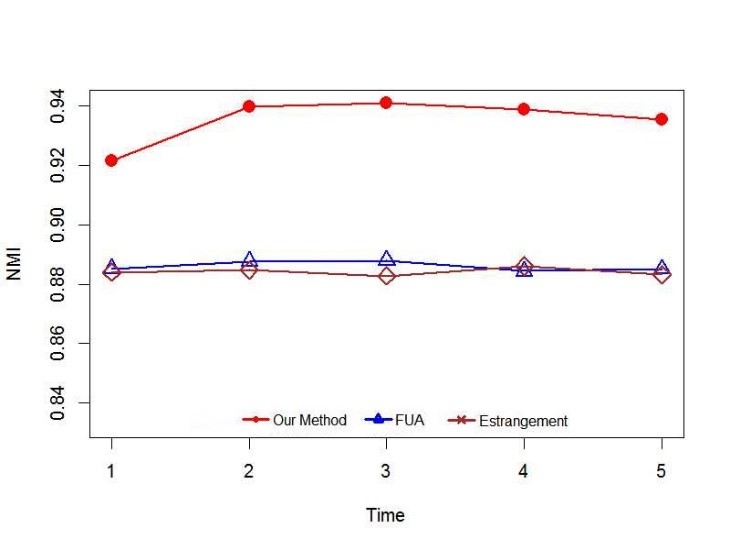}\par 
    \subcaption{Expansion and contraction events}
    
    \end{multicols}
\begin{multicols}{2}
    \includegraphics[width=\linewidth]{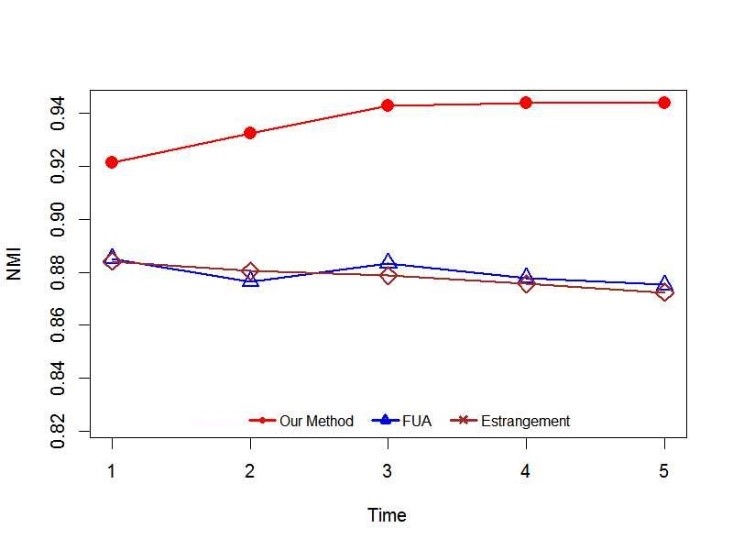}\par
    \subcaption{Birth and death events}
    \includegraphics[width=\linewidth]{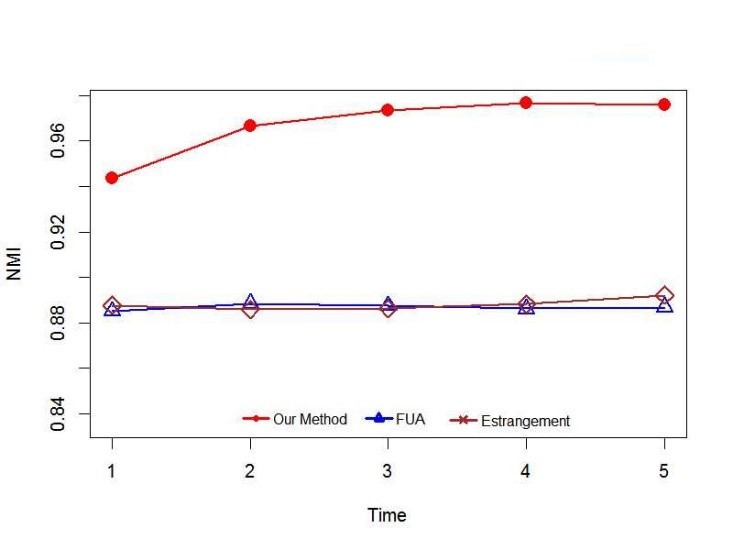}\par
    \subcaption{Merging and splitting events}
    
\end{multicols}

\begin{center}
\includegraphics[scale=1]{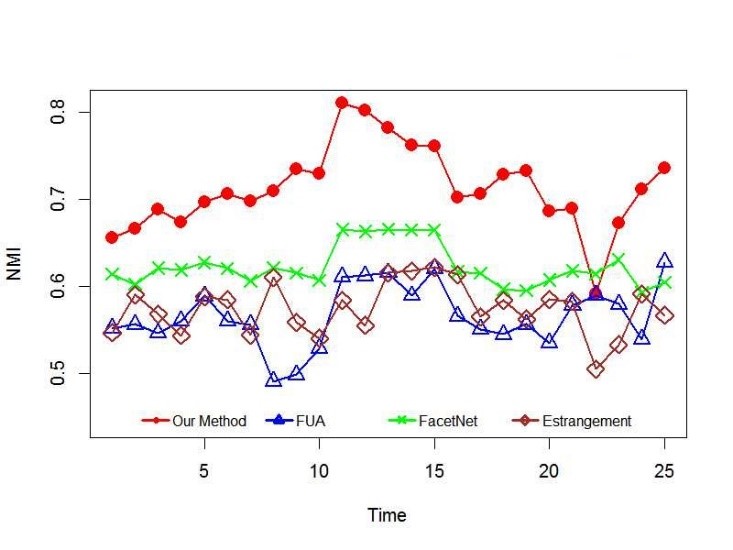}\par
    \subcaption{Kawadia synthetic network with  $n=100$, $P=0.4$, $P_c=0.2$, $P_r=0.05$ and $n_c=20$}
\end{center}
\caption{Similarity to groun-truth community in synthetic networks}
\label{fig6}
\end{figure*}

     After KIT is investigated, accuracy of our algorithm is tested on both sets of computer generated datasets (Fig.\ref{fig6}). The datasets provided by Green et al. aimed to capture four main events occurred during lifetime of a community. The results state that the accuracy of the proposed method increases, while the performance of other algorithms remains almost constant. For Kawadia synthetic network, our detected community structure is the most similar to the ground-truth community structure as well.

\subsection{Comparing the runtimes}
In this section we will compare the running times of dynamic clustering methods. The results are listed in table\ref{table1}.

The results of Facebook and Enron datasets on which Estrangement method was applied did not converge in 24 hours of runtime, thus they are omitt ed in table \ref{table1}. Moreover, the runtime results of these two datasets are ignored for FacetNet method since the corresponding process required more memory than the amount available in the test machine. It is obvious that the proposed method will be more scalable than the existing work, in confronting with big dynamic social networks.
\begin{table}[H]
\centering
\caption{Running time of different dynamic clustering methods}
\label{table1}
\begin{tabular}{cccccc}
\hline
                & Cellphone & Opsahl  & KIT      & Facebook & Enron  \\ \hline
FacetNet        & 8 sec     & 72 sec  & 163 sec  & -        & -      \\
Estrangement    & 543 sec   & 612 sec & 1532 sec & -        & -      \\
Proposed method & 33 sec    & 14 sec  & 210 sec  & 32 min   & 36 min \\ \hline
\end{tabular}
\end{table}

\section{Conclusion}
Detecting of communities and tracking their evolution in dynamic social networks is a significant research problem. In this work, we proposed a fast parameter-free method to find meaningful communities in highly dynamic social networks. By the concept of leadership, our algorithm first identifies some influential nodes called leaders and then employs these central members to expand local communities around them. Thanks to the incremental approach, our method reveals smooth community structure over time. We have conducted comprehensive experiments on both computer generated networks and real-world dynamic ones. Since our method employs the community structure of the previous snapshots, it seizes the community evolution over time. The obtained results, distinguish our algorithm from the existing dynamic clustering methods.
     The behavior of the community leaders is of a great importance in social networks. For future works, we plan to investigate the lifetime evolution of these promising members in various dynamic social networks. This information will be used to capture the main characteristics of such networks and enables us to predict the incoming structure of networks.

\end{document}